\documentclass[twocolumn,amsmath,amssymb,hidelinks]{revtex4}
\usepackage{CJKutf8}
\usepackage{graphicx}
\usepackage{color}
\usepackage{booktabs}
\usepackage{multirow}
\usepackage{enumitem}
\usepackage{hyperref}
\usepackage{xcolor}

\begin{document}
\title{General Mechanism of Evolution Shared by Proteins and Words}
\author{Li-Min Wang$^1$, Hsing-Yi Lai$^{1, \dagger}$, Sun-Ting Tsai$^{2, \ddagger}$, Chen Siang Ng$^3$, Kevin Sheng-Kai Ma$^4$, Shan-Jyun Wu$^1$,\\
 Meng-Xue Tsai$^1$, Yi-Ching Su$^5$, Daw-Wei Wang$^1$, and Tzay-Ming Hong$^{1,\ast}$}
\affiliation{$^1$Department of Physics, National Tsing Hua University, Hsinchu 30013, Taiwan, Republic of China\\
	$^2$ Department of Physics and Institute for Physical Science and Technology, University of Maryland, College Park, MD 20742, U.S.A.\\
	$^3$Institute of Molecular and Cellular Biology \& Department of Life Science, National Tsing Hua University, Hsinchu 30013, Taiwan, Republic of China\\
	$^4$Center for Global Health, Perelman School of Medicine, University of Pennsylvania, Philadelphia, Pennsylvania, U.S.A.\\
	$^5$Department of Linguistics, National Tsing Hua University, Hsinchu 30013, Taiwan, Republic of China
	}
\date{\today}

\begin{abstract}
Complex systems, such as life and languages, are governed by principles of evolution. The analogy and comparison between biology and linguistics\cite{alphafold2, RoseTTAFold, lang_virus, cell language, faculty1, language of gene, Protein linguistics, dictionary, Grammar of pro_dom, complexity, genomics_nlp, InterPro, language modeling, Protein language modeling} provide a computational foundation for characterizing and analyzing protein sequences, human corpora, and their evolution. However, no general mathematical formula has been proposed so far to illuminate the origin of quantitative hallmarks shared by life and language. Here we show several new statistical relationships shared by proteins and words, which inspire us to establish a general mechanism of evolution with explicit formulations that can incorporate both old and new characteristics. We found natural selection can be quantified via the entropic formulation by the principle of least effort to determine the sequence variation that survives in evolution. Besides, the origin of power law behavior and how changes in the environment stimulate the emergence of new proteins and words can also be explained via the introduction of function connection network. Our results demonstrate not only the correspondence between genetics and linguistics over their different hierarchies but also new fundamental physical properties for the evolution of complex adaptive systems. We anticipate our statistical tests can function as quantitative criteria to examine whether an evolution theory of sequence is consistent with the regularity of real data. In the meantime, their correspondence broadens the bridge to exchange existing knowledge, spurs new interpretations, and opens Pandora's box to release several potentially revolutionary challenges. For example, does linguistic arbitrariness conflict with the dogma that structure determines function?
\end{abstract}

\maketitle
\begin{CJK}{UTF8}{bsmi}
Understanding the universal characteristics of nature is one of the central problems in complex system sciences\cite{scale_inv, self-org, Tao, complex network, uni_network, social dynamics, music}, such as power-law behavior, hierarchical organization, and diversification. These characteristics motivate scientists to pursue an ultimate goal: a unified theoretical framework or general mechanism for understanding various phenomena in different systems. Here, as a small step toward this goal, we establish a common mechanism that underlies two important complex systems: biology and linguistics.

Life and language share similar hallmarks. In academic terms, they are both sequential information arranged hierarchically with discrete and unblendable units\cite{faculty1}, being heritable, and obeying Zipf’s law\cite{cell language, language of gene, Protein linguistics, dictionary, Grammar of pro_dom}. The hierarchy of life can be structured as amino acid $\rightarrow$ protein domain $\rightarrow$ protein $\rightarrow$ ... higher level; while in language, it can be phoneme $\rightarrow$ syllable $\rightarrow$ word $\rightarrow$ ... higher level. By combining small units into large units, the functions of a complex adaptive system are constructed. The diversification, i.e., simple to complex, is conditioned by evolution. Only a small fraction of all possible combinations is functional and survives in evolution, namely, the combination of units is not purely random. Then it rises a curial question: what is the underlying mechanism that determines the rule of combination?

Finding the analogy and comparison between biology and linguistics may act as the Rosetta Stone to decipher the language of life\cite{lang_virus, cell language, language of gene, Protein linguistics, Grammar of pro_dom, dictionary}. Notable examples of bioinformatic techniques that are grounded in linguistics and natural language processing are\cite{Grammar of pro_dom, dictionary, alphafold2, RoseTTAFold, lang_virus, complexity, genomics_nlp, InterPro, language modeling, Protein language modeling} 
({\rm i}) AlphaFold2 and RoseTTAFold which leverage multi-sequence alignments to predict protein structure from the amino acid sequence\cite{alphafold2, RoseTTAFold},
({\rm ii}) a neural language model that predicts viral escape\cite{lang_virus}, 
({\rm iii}) the application of rules of language to describe the organization and evolution of protein and its domains\cite{language of gene, cell language, Protein linguistics, Grammar of pro_dom}, 
and ({\rm iv}) probabilistic segmentation model to identify presumptive regulatory sites\cite{dictionary}. Reciprocally, linguists have also been inspired by biology to discover the secret of human language\cite{nl_ns, learn, faculty1, faculty2, lang_punctuation, social dynamics}. Famous instances include 
({\rm i}) the natural selection in languages\cite{nl_ns, learn}, 
({\rm ii}) the discussion of linguistic universal from the viewpoint of biolinguistics\cite{faculty1, faculty2, social dynamics}, 
({\rm iii}) the discovery that languages exhibit the signature of both gradual\cite{nl_ns} and punctuational evolution\cite{lang_punctuation}.

Nevertheless, the analogy between biology and linguistics is still highly speculative. In biology, most researchers either merely apply linguistic techniques or just qualitatively describe their relationship. In linguistics, most studies of linguistic universal concern the relationship between words. On the other hand, a general quantitative mechanism of word formation has never been found. Both biology and linguistics stop short of providing common formulations with rigorous evidence. To obtain a general mechanism, we begin with the determination of the correspondence between genetics and linguistics\cite{cell language, Protein linguistics} (GLC) over different hierarchies.

\begin{table*}[!t]
	\caption{Hierarchy for GLC is supported by qualitative or quantitative features. The GLC in this paper mainly focuses on the evolutionary mechanism for blocks and components. The function of a block is defined by its interaction with the environment. See METHOD for details and examples.}
\begin{tabular}{| c || c | c || l |}
\hline
{\bf GLC Hierarchy} & {\bf Life}                                                                             & {\bf Language}                                                                     & {\bf Common Features}                                                                                              \\ \hline\hline
Element                & standard amino acid                                                                            & phoneme                                                                       & size of the set is one order of magnitude                                                                        \\ \hline
Component              & domain                                                                           & syllagram                                                                      & $\rho_y$, Eqs. (\ref{scaling relation}, \ref{Allo_y}), $P(d_c)$                             \\ \hline
Block                  & protein                                                                          & word                                                                          & $\rho_x$, Eqs. (\ref{scaling relation}, \ref{Chain_x}), $P(d_b)$                             \\ \hline
Individual             & organism                                                                         & person (speaker)                                                                        & need functions to survive/communicate                                                                              \\ \hline
Environment              & ecological niche & community & \begin{tabular}[c]{@{}l@{}}affect the function connections in hierarchy\end{tabular} \\ \hline
\end{tabular}
	\label{hierarchy}
\end{table*}

\begin{figure*}[t!]
	\includegraphics[width=14 cm]{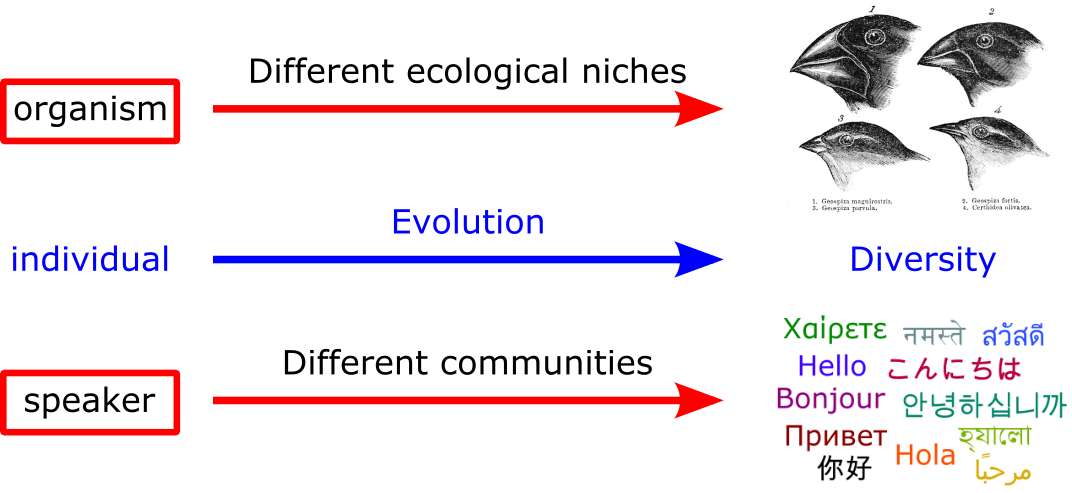}  
	\caption
	{Schematic of the common evolution framework according to GLC and Tab. \ref{hierarchy}. 
	}\label{framework}
\end{figure*}

\section*{Genetics-Linguistics correspondence}
The diversification of complex systems is governed by principles of evolution. An intuitive way to establish GLC starts from the macroscopic scale. To survive in different environments, an individual will evolve and lead to diversity. As shown in Fig. \ref{framework}, the relation (individual, environment) becomes (organism, ecological niche) for life and (speaker, community) for language. In fact, the framework of GLC provides an alternative viewpoint on language evolution which comes from the interaction between individuals and environment, as opposed to between two individuals in communication model\cite{social dynamics}. Note that the relation of  (organism, ecological niche) in life does not correspond with that of (speaker, another speaker) in language within the framework of communication model. See METHOD for how the relation (individual, environment) affects the interpretation of mathematical formulations.

After establishing the macroscopic GLC, let us build the microscopic framework on  heritably functional units which are subjected to selection. For genetics, we analyze the amino acid sequence (AAS). It is because once there is information on AAS, the structure of a protein can be found\cite{alphafold2, RoseTTAFold}, which conditions the molecular functions. While for linguistics, we focus on multilingual corpora since they are ideal for quantitative analysis. At the same time, it is necessary to find the linguistic universal shared by spoken and written languages. So we will consider the unit of spoken form, then generalize such a unit to written form. See Supplementary Information, SI, for data construction and the role of gene in the first GLC.

The first GLC is standard amino acid $\leftrightarrow$ phoneme as described in Tab. \ref{hierarchy}, where ``$\leftrightarrow$" denotes ``corresponding to". It is based on the fact that both AAS and corpora are formed by a set of finite elements whose size (cardinality) is about one order of magnitude. The AAS is formed by 20 kinds of standard amino acids, while for world languages\cite{world lang}, such as English and Mandarin (Chinese), the cardinalities of their phonemic inventories are also one order of magnitude\cite{phone-inven}. See SI for more discussions.

\begin{figure*}[h!]
	\includegraphics[width=16cm]{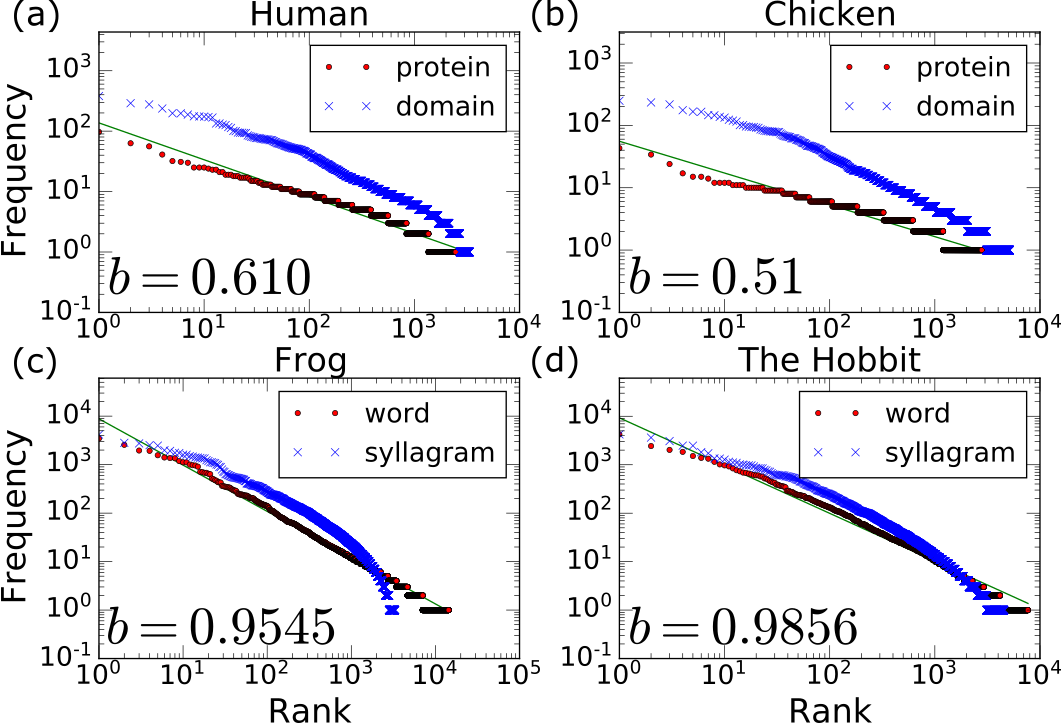}  
	\caption
	{ The dotted/crossed (red/blue) curves in the log-log plots of (a, b) show FRD of protein/domain in the genome of human and chicken. Similarly, (c, d) are for word/syllagram in the Mandarin novel, Frog, and the English novel, The Hobbit. We fit the data with maximum likelihood estimate\cite{AIC}. The fitting function is for the red points.
	}\label{FRD}
\end{figure*}

The second GLC is domain $\leftrightarrow$ syllagram. The latter is a newly introduced linguistic unit which will be defined in the next paragraph. Identification of this relation takes more evidence to substantiate. Intuitively, it comes from the combination of elements. In genetics, domain\cite{Protein linguistics, multi_dom, pro_univ and gene_evo}, i.e., the combination of amino acids, is a functional and evolutionary unit of proteins. Some researchers believe that domain is the ``word of gene"\cite{language of gene, Grammar of pro_dom, lang_pro_univ} because its frequency-rank distribution (FRD) seems to obey an important statistical property for word: Zipf's law\cite{N-Gram Categorization, Zipf} $\rho_x=a/x^b$ where $\rho_x$ denotes the frequency of occurrence of word whose rank is $x$, and $b\approx1$. However, a simple and strong reason can refute this analogy. Unlike a sentence, which is rarely repeated in a piece of writing, a protein usually appears multiple times in AAS. So protein does not function as the sentence, neither does domain the word. As shown in Extended Fig. \ref{FRD}, both FRD of protein and word $\rho_x$ fit power law, while those of domain and syllagram $\rho_y$ follow a similar curve. This fact seems to imply that protein $\leftrightarrow$ word is the third GLC.

In linguistics, syllable, namely, the combination of phonemes, is the phonological unit of words. If the second GLC is built on syllable, there is a problem: it can only be defined in spoken form\cite{syl-def}. To solve this issue, we define a new linguistic unit, syllagram, that can be a bridge connecting speaking and writing. A syllagram is defined as a unit in written form which represents the corresponding syllables in a word. For example in English, the syllagram ``sy-" in ``syllable" and ``si-" in ``silly" share the same pronunciation but are spelled by different letters. Similar examples can also be found in Mandarin. For instance, the syllagrams ``敢" (dare) in ``勇敢" and ``趕" (hurry) in ``趕車" have the same pronunciation [g\v an] but are of different written forms. See SI for more details about syllagram. As in Extended Fig. \ref{FRD}, the FRD of syllagram is different from that of word, which seems to assume that the second GLC would be domain $\leftrightarrow$ syllagram. 

In fact, it is not sufficient to uphold such assumptions for the second and third GLC because we have not checked how similar is the relation between domain and protein to that between syllagram and word, which cannot be manifested via FRD. In the following sections, we are going to verify our assumption upon closer scrutiny.

\section*{Rank-Rank Analysis and Scaling Structure}
\begin{figure}[h!]
	\includegraphics[width=8cm]{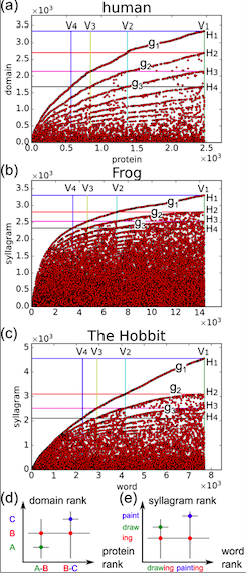}  
	\caption
	{ Panels (a, b, c) are the RRD plot for Human, the novels Frog, and The Hobbit. Their construction is demonstrated schematically in panels (d, e). The introduction of vertical $V_m$ and horizontal lines $H_n$ is instrumental to facilitate the understanding of scaling structure. See SI for more details.
	}\label{RRD}
\end{figure}

A key to verifying our assumption lies in the rank-rank distribution (RRD) which is used to graphically demonstrate how proteins/words are composed of domains/syllagrams. The prerequisite (see METHOD) of constructing RRD is to segment (protein, domain) from genome\cite{InterPro, Ensembl}, and (word, syllagram) from corpora (see SI).  Now let us build the relationship between unit and subunit. For genetic data, let $(x,y)=($rank of protein, rank of domain$)$; while for linguistic data, $(x,y)=($rank of word, rank of syllagram$)$, where the ranks are determined by FRD $(\rho_x, \rho_y)$. Figure \ref{RRD} exhibits the RRD in (a) the human genome, (b, c) Mandarin and English novels, while (d, e) exemplify the process of plotting RRD. 

\begin{figure*}[]
	\includegraphics[width=16cm]{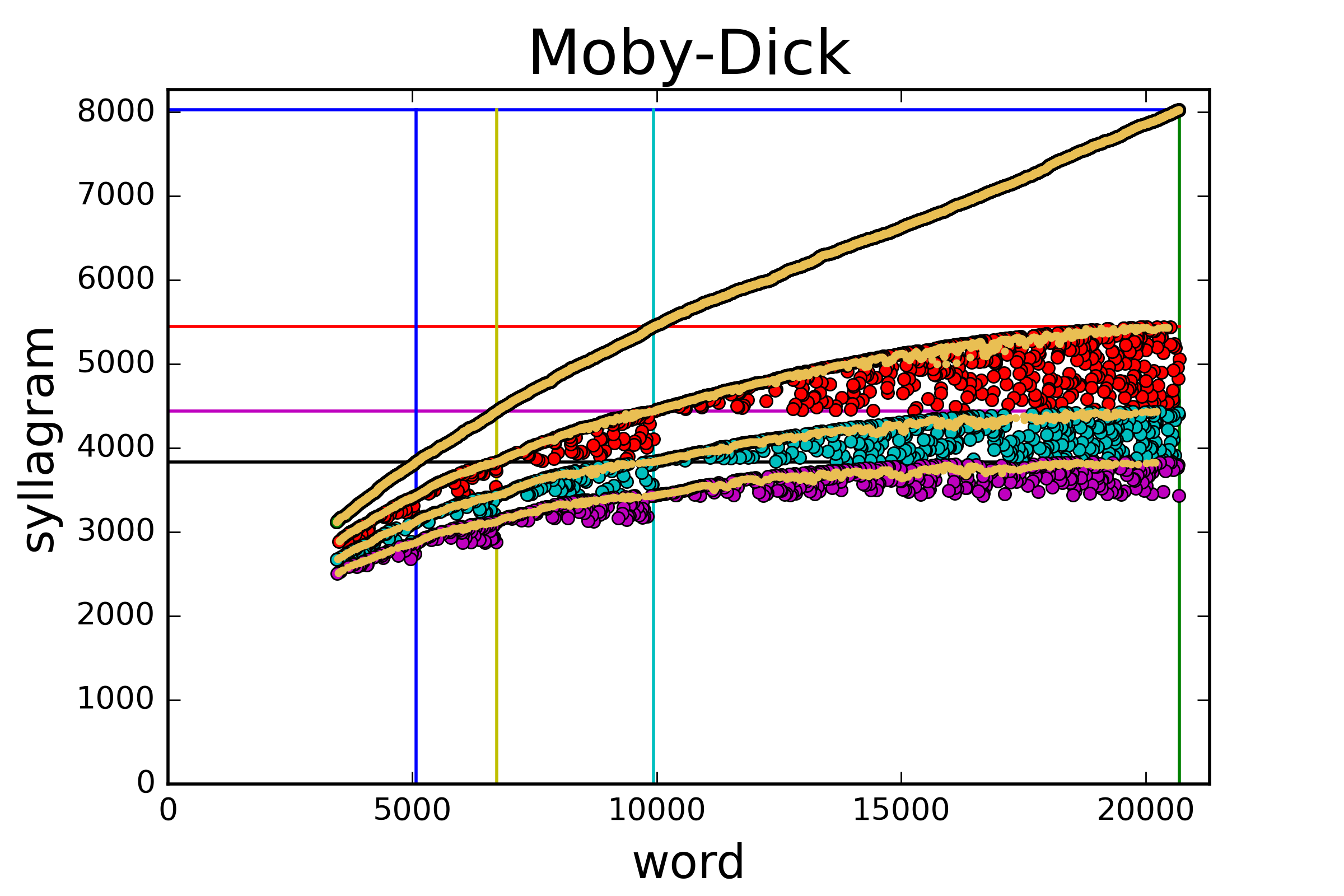}  
	\caption
	{Our de-noise algorithm can sketch out the scaling structure of RRD, as yellow points. Here we exemplify it for the English novel, Moby-Dick. The red, green, and purple points are ``fog" of $g_2, g_3,$ and $g_4$. See SI for technical details.
	}\label{denoise}
\end{figure*}

\begin{figure*}[]
	\includegraphics[width=16cm]{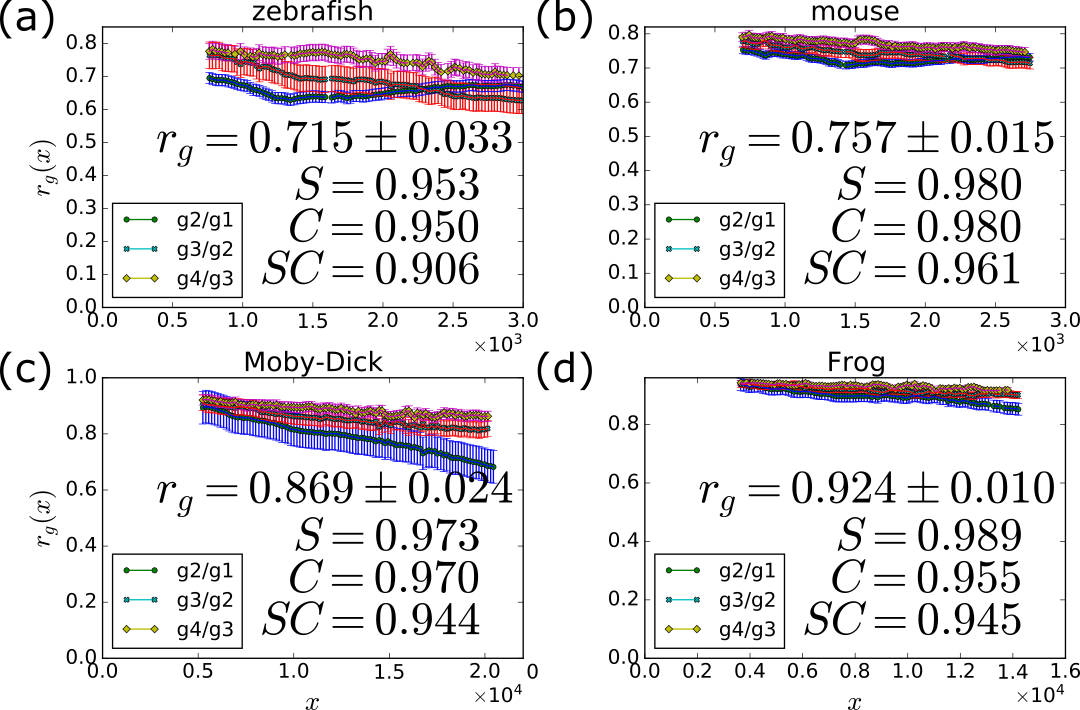}  
	\caption
	{These figures depict $r_g$ with standard error $\sigma_g$. Panel (a) is for the zebrafish genome, (b) for the mouse genome, (c) for the English novel {\it Moby-Dick}, and (d) for the Mandarin novel {\it Frog}. Rigorous definitions of $r_g$ and $SC$ value, a measure of goodness of scaling, are given in SI. Note that we do not consider "$g_2/g_1$" when calculating $SC$. See SI for technical details and reasons.
	}\label{scaling}
\end{figure*}

After inspecting RRD for the genomes of 201 organisms and 67 corpora (see SI for data), we find a universal phenomenon: the envelopes that comprise this structure, labeled as $g_\ell$ where $\ell$ denotes the $\ell$-th curve (see the yellow curves in Extended Fig. \ref{denoise}), obey a scaling relation: 
\begin{equation}
g_{\ell+1}(x)/g_\ell(x)=r_g
\label{scaling relation}\end{equation}
where $r_g$ is a constant of $\ell$ and $x$, as verified in Extended Fig. \ref{scaling}. Besides, the soundness-clearness value (SC value, see SI for definition), an index to describe the goodness of scaling, is larger than 0.7 for most of our data. Thus, the scaling structure is evidence for the second and third GLC. We conjecture such a structure generally exists in different organisms and human languages. One cannot help but marvel at this structure which is not accidental. See SI for further discussion.

In addition to $\rho_x$, $\rho_y$, Eq. (\ref{scaling relation}), and $SC$ value, there are two other properties hidden behind RRD. We realize the data points $\vec{r}=(r_b, r_c)$ on RRD can be grouped into
\begin{equation}
\begin{aligned}
D_y(R_c)\equiv\{\vec{r}=(r_b, r_c)\ \vert\ r_c = R_c\}\\
D_x(R_b)\equiv\{\vec{r}=(r_b, r_c)\ \vert\ r_b = R_b\}
\end{aligned}
\end{equation}
where $\vec{R}=(R_b, R_c)$ is the selected point. The set $D_y(R_c)$ indicates the words composed of the syllagram with rank $R_c$, while $D_x(R_b)$ indicates the syllagrams used in the word with rank $R_b$. The two hidden properties can be defined as
\begin{equation}
Allo(R_c)\equiv|D_y(R_c)|
\label{Allo}\end{equation}
\begin{equation}
Chain(R_b)\equiv\sum_{\vec{r}\in D_x(R_b)} Allo(r_c)
\label{Chain}\end{equation}
for $\vec{R}=(R_b, R_c)$ and $\vec{r}=(r_b, r_c)$, where $|D_y(R_c)|$ denotes the cardinality of $D_y(R_c)$. The allocation function $Allo(R_c)$ represents the ability to allocate a syllagram $R_c$ to other words, while the chain function $Chain(R_b)$ indicates how a word $R_b$ links to other words. See SI for examples.

\begin{figure*}[]
	\includegraphics[width=16cm]{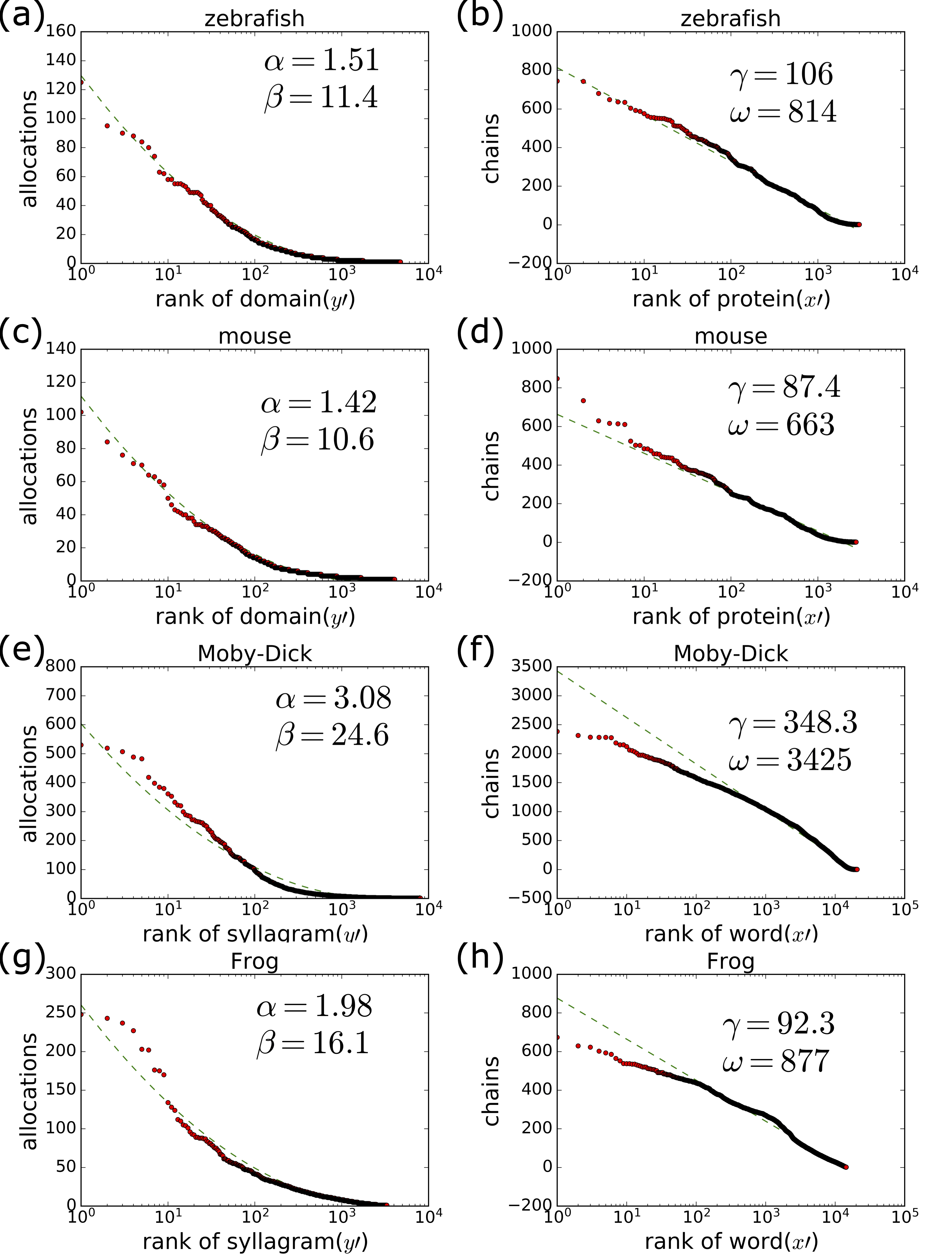}  
	\caption
	{Left figures depict $Allo$-rank plot, while right ones are for $Chain$-rank. Panels (a, b) are for the zebrafish genome, (c, d) for the mouse genome, (e, f) for English novel {\it Moby-Dick}, (g, h) for Mandarin novel {\it Frog}. Dashed lines are fitting curves, while fitting parameters $\alpha, \beta. \gamma$, and $\omega$ are defined in Eq. (\ref{Allo_y}, \ref{Chain_x}) and fitted via non-linear least squares. See SI for further information.
	}\label{LC}
\end{figure*}

By fitting real data for genomes and corpora, as in Extended Fig. \ref{LC}, we observe that they satisfy two simple empirical relations:
\begin{equation}
Allo(y')=(-\alpha\ln{y'}+\beta)^2
\label{Allo_y}\end{equation}
\begin{equation}
Chain(x')=-\gamma\ln{x'}+\omega
\label{Chain_x}\end{equation}
where $\vec{R}'=(x', y')$ is the new rank-rank vector depending on $(Chain, Allo)$ instead of $(\rho_x, \rho_y)$, and $\alpha, \beta, \gamma, \omega > 0$ are fitting parameters (see SI).

The $Chain$ and $Allo$ functions unveil the hidden interrelationship between protein/word and domain/syllagram, while $\rho_x$ and $\rho_y$ present their individual relationships. Combining these quantitative hallmarks, we provide more evidence to support our assumption that domain $\leftrightarrow$ syllagram and protein $\leftrightarrow$ word are the second and third GLC, respectively. In the next section, we will show further features of $D_x$ and $D_y$.

\section*{Network Analysis}
\begin{figure*}[h!]
	\includegraphics[width=16cm]{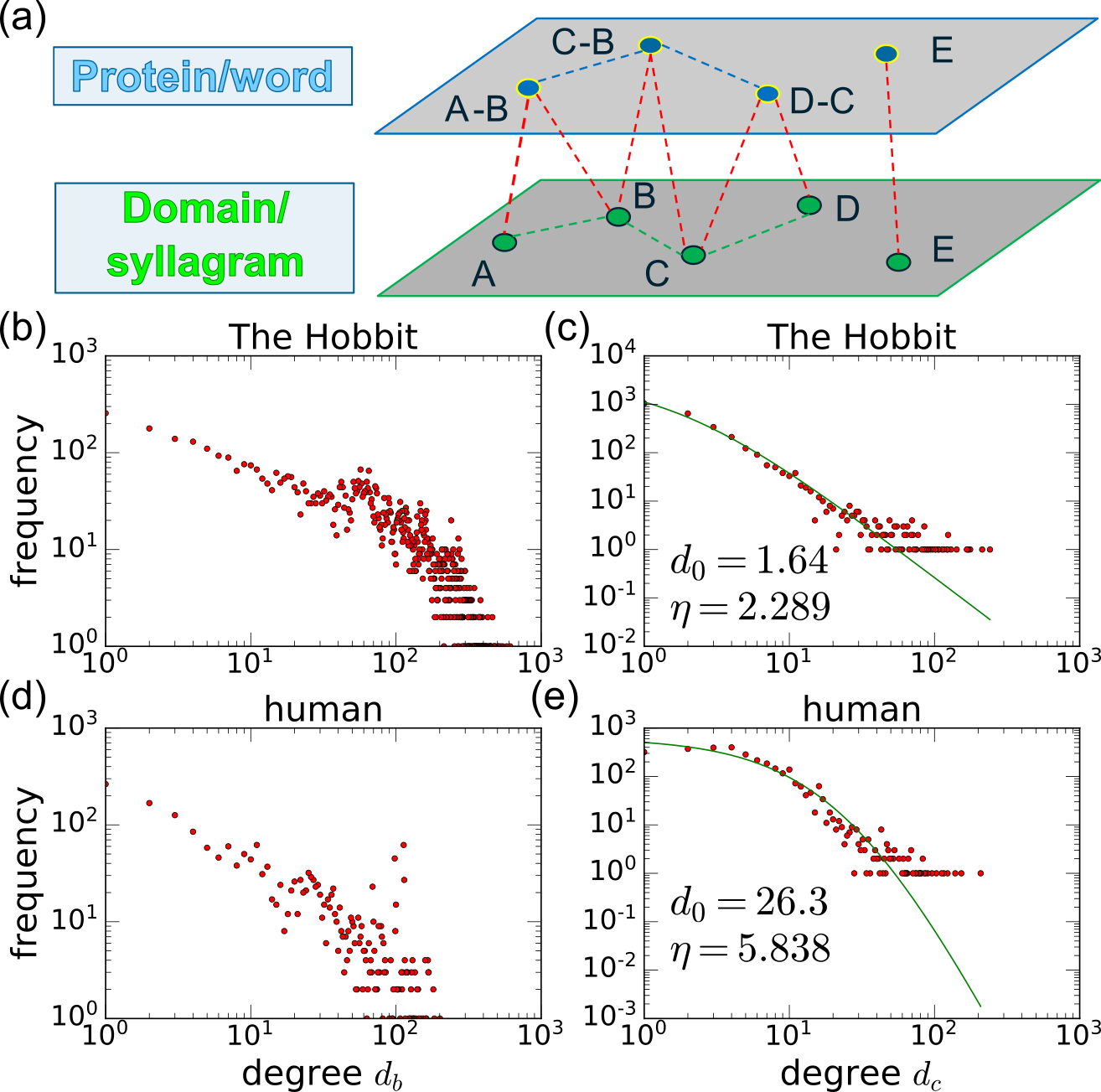}  
	\caption
	{The schematic of multilayer network for protein/word and domain/syllagram layers is in panel (a). Panels (b, c) show the degree distribution of (word, syllagram) in The Hobbit, while (d, e) show that of (protein, domain) in the human genome. The vertices with zero degrees have been excluded because they are presented as singularities. Note that (c, e) exhibit the feature of a shifted power law network\cite{shifted} $P(d_c)\sim (d_c+d_0)^{-\eta}$. We fit the data with maximum likelihood estimate\cite{AIC}, see SI for details.
	}\label{topology}
\end{figure*}

Network is a handy tool to describe the dynamical processes in complex systems\cite{scale_inv, uni_network, MN, complex network, BA model, SF in domain network, self-org, shifted, social dynamics}. We realize that $D_x$ and $D_y$ can be used to construct a multilayer network that contains two layers, protein/word $G_b$ and domain/syllagram $G_c$, as in Extended Fig. \ref{topology}. Whenever two words/syllagrams appear in the same $D_y$/$D_x$, an edge $e^b$/$e^c$ is assigned between them. Same for proteins/domains.

We checked that the features in Extended Fig. \ref{topology} are shared by different genomes and corpora (see SI). The degree distribution $P(d_\theta)$ denotes the number of vertices exhibiting $d_\theta$ edges, where $\theta=b$ for protein/word and $c$ for domain/syllagram. The shifted power law\cite{shifted} behavior we discovered for $G_c$ not only agrees with previous research on protein sequences\cite{SF in domain network} but also exists in languages (see Supplementary Data). 

With the aid of rank-rank and network analyses, we found many quantitative hallmarks shared by AAS and corpora: ({\rm i})  frequency-rank distribution of protein/word obeys power law, ({\rm ii})  rank-rank distribution exhibits the scaling structure, ({\rm iii})  allocation and chain function, as expressed by Eqs. (\ref{Allo_y}, \ref{Chain_x}), are new properties, and ({\rm iv}) network of domain/syllagram obeys shifted power law. So far, these facts strongly uphold the existence of GLC in Tab. \ref{hierarchy}. To build a common framework, we give collective nouns to different hierarchies of life and language. Now a critical question emerges: how to establish a general mechanism to reproduce all characteristics mentioned above for life and language? In the next section, we will show the key to answering these questions.

\section*{Block-Function Association and\\ Function Connection}
\begin{figure}[h!]
	\includegraphics[width=8cm]{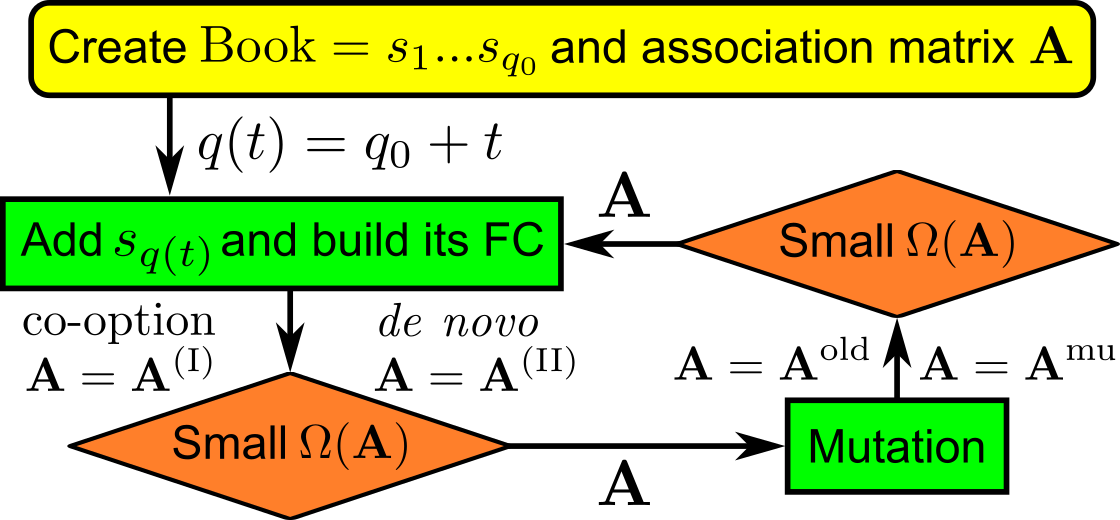}  
	\caption
	{Simple flowchart of the evolutionary algorithm in our mechanism of evolution. The Book denotes a sequence of proteins/words, as in Eq. (\ref{Book}). The step ``Add $s_{q(t)}$ and build its FC" is the start of loop. It simulates the sequence variation that changes the length of Book $q(t)$. This loop will execute from $t=1$ until $t=L-q_0$ where $L$ is the final length of Book. The step ``Small $\Omega({\bf A})$" uses the principle of least effort to simulate natural selection. The step ``Mutation" simulates the sequence variation which does not change $q(t)$. See METHOD for details and SI for the fast algorithm.
	}\label{flowchart}
\end{figure}

Both life and language would undergo the heritable diversification. To simulate AAS and corpora, our mechanism generates a sequence of blocks. Several ``sequence variations" will be produced to simulate the genetic and linguistic variations. Then a quantitative version\cite{origin_Zipf} of the principle of least effort\cite{Zipf} will be used to simulate the natural selection and determine whether a sequence variation is beneficial for survival/communication. The whole mechanism can be simulated through an evolutionary algorithm as in Fig. \ref{flowchart}.

Now let us introduce our mechanism. Assume there are $p$ blocks $\mathcal{B}=\{b_1, ...,b_p\}$. Since both AAS and corpora are sequential information, they can be written as ``Book":
\begin{equation}
{\rm Book}(b_1, ...,b_p) = s_1s_2...s_q
\label{Book}
\end{equation}
where $\text{Book}: \mathcal{B}^p\to\mathcal{B}^q$ and $s_1, ..., s_q\in\mathcal{B}$. For each block $s_j$, we denote its function, which is determined by the interaction between blocks and environment, as $f_j$. So that the functional representation of ``Book" is
\begin{equation}
Fr({\rm Book}) = f_1f_2....f_q
\label{Fr_Book}\end{equation}
that acts for the functions $\mathcal{F}=\{f_1, ...,f_q\}$ for an individual (survival or communication\cite{nature affect0, nature affect1, phonology}). The mapping $Fr: \mathcal{B}^q \to\mathcal{F}^q$ can specify the functions of Book. Combining Eqs. (\ref{Book}, \ref{Fr_Book}), the function composition $Fr\circ\text{Book}$ can be described by an association matrix ${\bf A}:\mathcal{B}^p \to\mathcal{F}^q$ so-called block-function association (See METHOD). 

To ``write" such a functional Book, the individual needs to pay two kinds of effort. The first is ``individual effort" $\mathcal{H}_p(\mathcal{B})$. The fewer varieties of $\mathcal{B}$ are, the less effort the individual needs to pay. This prompts the individual to produce fewer kinds of blocks. The second is ``collective effort" $\mathcal{H}_q(\mathcal{F} | \mathcal{B})$. Once a block is produced, it is used to carry out some functions. If the block provides only a few functions, it can work more specifically and therefore decrease the collective effort for the individual. 

\begin{figure}[h!]
	\includegraphics[width=8.1cm]{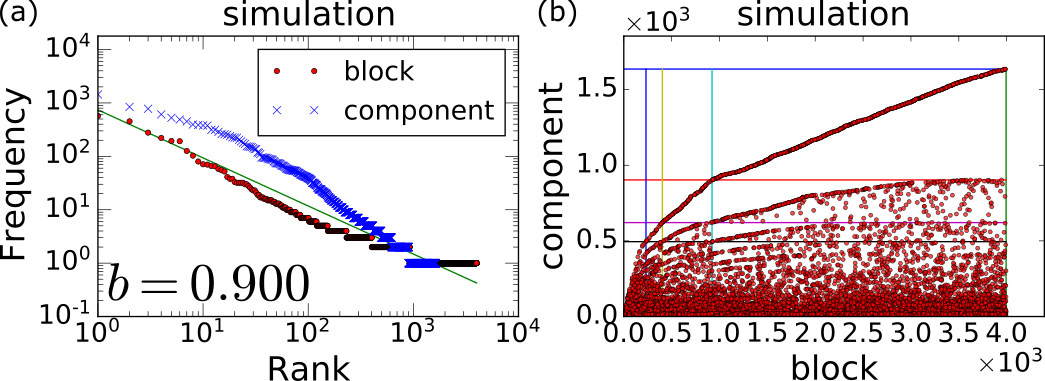}  
	\caption
	{(a) FRD and (b) RRD of our simulation exhibit scaling structure. The exponent $b$ of FRD $\rho_x$ can be varied by adjusting the mutation rate $P_{\text{mu}}$. As $b = 0.4\sim 0.7$, the simulation behaves like life; while $b =  0.8 \sim 1$, it behaves like language. See METHOD and SI for details, and Extended Fig. \ref{more_simulation} for parameters and other characteristics.
	}\label{simulation}
\end{figure}

The cost function of total effort is
\begin{equation}
\Omega_\lambda({\bf A})=\lambda \mathcal{H}_q(\mathcal{F} | \mathcal{B})+(1-\lambda)\mathcal{H}_p(\mathcal{B})
\label{tot_effort}\end{equation}
where $0\le\lambda\le1$ is a predefined parameter decided by the system. When a sequence variation happens, the association matrix ${\bf A}$ will be changed and revised $\Omega_\lambda({\bf A})$. Comparing $\Omega_\lambda({\bf A}^{\text{new}})$ with $\Omega_\lambda({\bf A}^{\text{old}})$, we will pick out the smaller one since it has a better evolutionary advantage to survive/communicate according to the principle of least effort. In other words, a quantified criterion for natural selection is given by the principle of least effort. The above model will give power law for AAS and corpora (see METHOD for details), as shown in our simulation in Fig. \ref{simulation}(a) and the past research\cite{origin_Zipf}. 

Equation (\ref{tot_effort}) reveals two fundamental properties of a complex adaptive system: ({\rm i}) Unification which comes from the fact that the individual chooses a block randomly instead of based on its function. This property is consistent with an important insight in linguistics: the existence of arbitrariness, which refers to the random choice of a signifier in languages; ({\rm ii}) Diversification which originates from the specialization of blocks. Put differently, the molecular function in genetics is determined by the collective interaction between a protein and the ecological niche; while in linguistics, this property renders the convention of mapping from a signifier to signified hard to change. 

\begin{figure}[h!]
	\includegraphics[width=8cm]{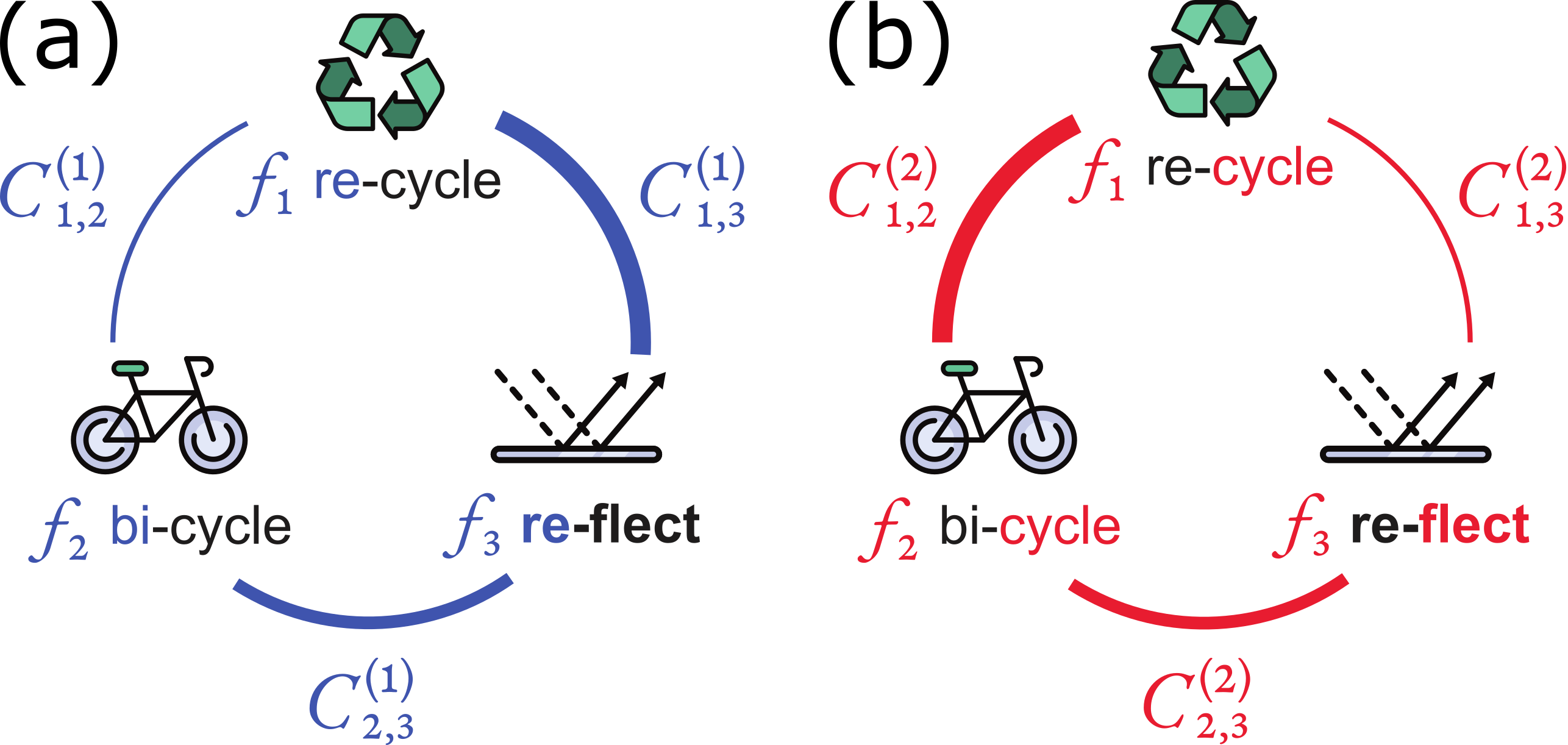}  
	\caption
	{The concepts, $f_1, f_2$, and $f_3$, are connected via their first syllagram in (a) or second one in (b). The thicker line indicates a stronger FC. When words for $f_1$ and $f_2$ already exist, we want to come up with a new word for $f_3$ which is more likely to adopt the syllagram from the old word that exhibits a stronger connection. Here, $f_3$ may be associated with ``reflect".
	}\label{logic_connect}
\end{figure}

Although the block-function association model successfully explains the origin of power law, it does not answer an essential question: how blocks are formed by their components?  To answer this, we introduce function connection (FC), which is defined as the correlation between functions. The stronger FC between two functions is, the higher the possibility that they get similar components. For instance, the proteins in the same family are composed of similar domains and so have correlated functions. The concepts of ``bi-cycle", ``re-cycle", and ``re-flect" are correlated by their syllagram. The process to determine FC is comprehensive (see METHOD) and will be influenced by the position of block in Eq. (\ref{Book}), similarities between functions, logical connections, cultures, etc. In other words, the interaction between environment and blocks determines FC. As the evolutionary pressure mounts, the individual needs a new function $f_{q+1}$ to survive. There are two situations: ({\rm I}) conserved change, $f_{q+1}$ does not affect old FCs between $f_j\in\{f_1, ..., f_q\}$; ({\rm II}) radical change,  $f_{q+1}$ affects old FCs. To show the simplest case that exhibits the hallmarks of evolution, we will exemplify our model in ({\rm I}).

\begin{figure*}[t!]
	\includegraphics[width=16cm]{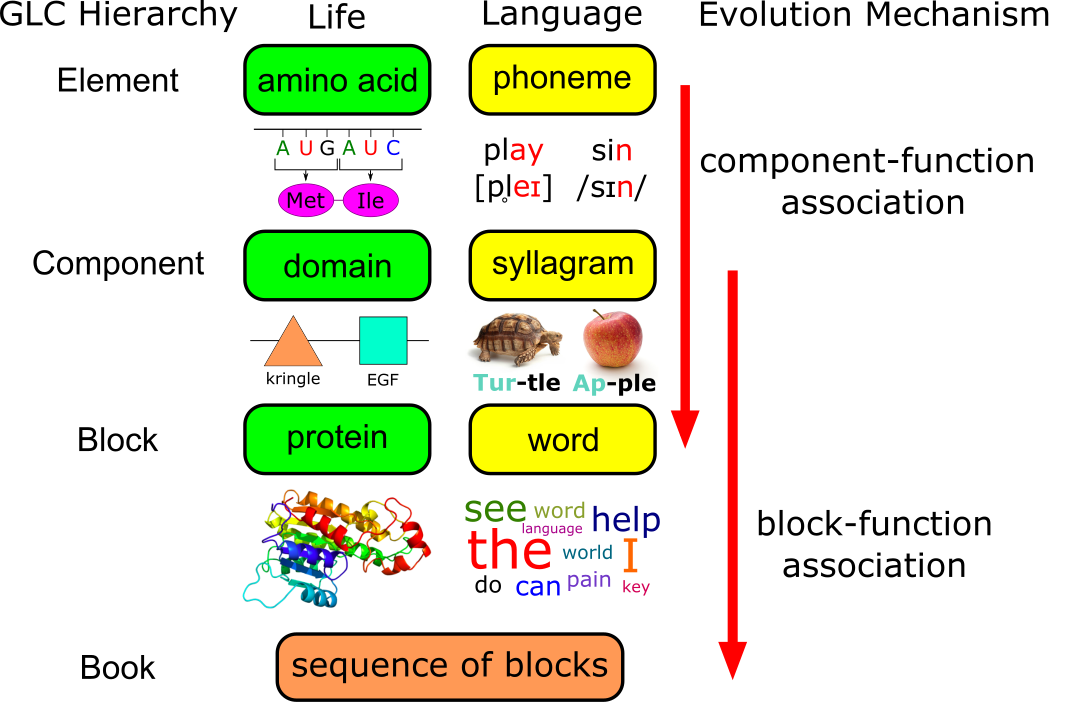}  
	\caption
	{A schematic that exhibits the evolution mechanism of different hierarchies. 
	}\label{Evo_hier}
\end{figure*}

The FC originates from components of which the position is crucial. For instance, there is ap-ple but no ple-ap in English. Another example can be found in order of domain combination\cite{AB not BA}. We denote the FC for the component at $k-$position between $f_{\mu}$ and $f_{\nu}$ as $C^{(k)}_{\mu,\nu}$ (see METHOD). Let us use Fig. \ref{logic_connect} to illustrate. Assume there were two existing words with the FCs for the first (blue) and second (red) syllagrams. Now the speaker wants to name a new concept $f_3$ with a 2-syllagram word. It would consist of either the old or new syllagram, so there are only two possible modes: ({\rm i}) co-option\cite{co-option}: assigning an old block $s_1$ or $s_2$ to the new function $f_3$, and ({\rm ii})  {\it de novo}: creating a new block that contains a new component to associate with the new function. If $f_3$ chooses ({\rm i}) co-option, its block may be either bicycle or recycle; ({\rm ii}) {\it de novo}, its block may be one of $\delta_1$-cycle, bi-$\delta_2$, re-$\delta_2$, or $\delta_1$-$\delta_2$ where $\delta_1\not\in$ \{bi, re\} and $\delta_2\not\in$ \{cycle\} denotes new components. Since both co-option and {\it de novo} are sequence variations, we can apply Eq. (\ref{tot_effort}) to determine which mode is better. The same model can also elucidate how a new protein is combined with domains. Via simulation (see Fig. \ref{flowchart} and Extended Fig. \ref{Evo_hier}), we can reproduce all the quantitative features listed in Tab. \ref{hierarchy} (see Fig. \ref{simulation}, Extended Fig. \ref{more_simulation}, and SI). See METHOD for mathematic details.

\section*{Component-Function Association Hypothesis}
In this section, we propose a ``component-function association" hypothesis as the evolution mechanism from the level of element to component. Recall that the first GLC is based on the fact that the cardinality of element inventory $O_1$ is one order of magnitude. For a component consisting of $L_c$ elements, the number of possible combinations is about $O_1^{L_c}$. Imitating Eq. (\ref{Book}) with $r$ components $\mathcal{C}=\{c_1, c_2, ..., c_r\}$, a block can be represented as the sequence of components
\begin{equation}
\text{Block}(c_1, ..., c_r) = I_1I_2...I_N
\label{Block}\end{equation}
where $\text{Block}: \mathcal{C}^r\to\mathcal{C}^N$ and $I_1, ..., I_N\in\mathcal{C}$. Similar to the FC network at component level in Extended Fig. \ref{FC_network}, the content of component is affected by the FC network at the element level. The Eq. (\ref{tot_effort}) from the principle of least effort can be used to decide the content of Block. 

Similar mathematical structures follow the same physical principle: a functional sequence is neither the simplest nor the most complex. However, there are some differences between block-function and component-function associations. First, the length of Book in Eq. (\ref{Book}) is much longer than that of Eq. (\ref{Block}). The former can be infinite, but the latter not. A general mechanism to decide the length of Block is still unknown. Second, the number of elements in a component seems not to be randomly decided. Take the phoneme $\langle a \rangle$ whose International Phonetic Alphabet number is 304 415 as an example. There is no syllable pronounced as $[aa]$ because the vocal structure of phoneme $\langle a \rangle$ provides a syllable boundary. In other words, the natural structure of elements forces the FC of some elements in certain positions to be always zero.

\section*{Conclusion}
The analogy between biology and linguistics has been touted as the Rosetta stone to decipher the language of life and human\cite{lang_virus, cell language, language of gene, Protein linguistics, Grammar of pro_dom, dictionary, nl_ns, learn, faculty1, faculty2, lang_punctuation, social dynamics}. In this paper, we have two major contributions. First, the establishment of genetics-linguistics correspondence as in Tab. \ref{hierarchy} reveals the quantitative characteristics shared by life and language (see SI and Supplementary Data for the analyses of 202 genomes of different organisms and 60 corpora of different languages). Several tools and independent statistical indices are proposed to describe the universal characteristics of life and language, such as organization between (protein, domain) and (word, syllagram), Zipf's law for frequency of occurrence, and shifted-power law in network. They can be functioned as quantitative criteria to examine whether an evolution theory of sequence is consistent with the regularity of real data. Second, a mechanism of evolution is shared by the sequences of protein and word, for which the finding of universal regularities helps elucidate the origin of molecular functions as well as human cognition. 

Our algorithm generates the sequence of proteins/words and simulates genetic/linguistic variations via the function connection networks. The entropic formulation quantifies the principle of least effort and natural selection and enables us to locate and explain the underlying mechanism for the origin of power law, the complexity of AAS/corpora, the composition of new protein/word, and all other characteristics of GLC (see METHOD and SI for details). Briefly, these feats are enabled by the fact that evolution obeys a fundamental principle: a functional sequence is neither the simplest nor the most complex. The framework of GLC not only interprets the universal properties of life and language, but also has the potential to be generalized to other complex systems with structural information, such as music\cite{language_music}. Additionally, the GLC framework can be used to construct interpretable machine learning models and correct the errors inside black box machine learning models. 
\end{CJK}

\section*{Author contributions}
T.M.H. and L.M.W. prepared the manuscript. L.M.W., H.Y.L., and T.M.H. proposed the theoretical models. L.M.W. designed the statistical indices and dealt with algorithms and technical details. S.T.T. assisted with the design of formulations, collected and tested corpora, and helped revise the manuscript. C.S.N. provided the knowledge of genome and protein sequences and helped revise the manuscript. S.J.W. and M.X.T. designed the basic function of program. Y.C.S. provided the knowledge of linguistics. D.W.W. helped explain the scaling structure.  
\section*{Competing interests}
The authors declare no competing interests.
\section*{Code availability}
Our codes are open and available in Wang, Li-Min \& Wu, Shan-Jyun. Genetics-Linguistics-Correspondence, Open source project on GitHub: \url{https://github.com/godofhoe/Genetics-Linguistics-Correspondence}

\section*{Acknowledgements}  We gratefully acknowledge useful discussions with Chih-Yu Yeh, Lee-Wei Yang, Chia-Hung Yang, Ching-Yao Lai, Tsung-Han Kuo, En-Jui Kuo, Mao-Syun Wong, Von-Wun Soo, Chi-Fang Chen, Zhengning Yang, and financial support from MoST in Taiwan under Grants No. 105-2112-M007-008-MY3 and No. 108-2112-M007-011-MY3.
\clearpage

\section*{METHODS}
\subsection{Prerequisite}
Before performing the rank-rank analysis, one needs to construct a text with segmented blocks and components. The details on how to construct a genome or linguistic data into Book can be found in SI. In this work, ({\rm i}) InterPro\cite{InterPro} is used to classify protein and domain from Ensembl\cite{Ensembl}, ({\rm ii}) several word segmentation systems\cite{Sinica, Standford1, Standford2} are employed for Mandarin texts, and ({\rm iii}) syllabipy\cite{syllabipy} is implemented to syllabify words from English texts.

\subsection{Block-function association}
Although the block-function association is a variant and generalization of the signal–object association\cite{origin_Zipf}, we introduced different assumptions and interpretations. Besides, the signal–object association model did not answer a fundamental problem: how does word formation originate? So we must elaborate on our model.

Considering a system with $p$ blocks $\mathcal{B}=\{b_1, ...,b_p\}$ and $q$ functions $\mathcal{F}=\{f_1, ...,f_q\}$. Each function is associated with some blocks and described by a binary matrix ${\bf A}=\{a_{ij}\}$ where $1\le i\le p$ and $1\le j \le q$. If the block $b_i$ refers to the function $f_j$, $a_{ij}=1$; otherwise $a_{ij}=0$. As long as a block is in use, it is assumed to exhibit certain functions irrespective of whether they have been identified. So the probability of producing $b_i$ is\cite{origin_Zipf}
\begin{equation}
P(b_i)=\sum_{j=1}^q P(b_i, f_j)
\label{P(b_i)}\end{equation}
where $P(b_i, f_j)$ is the joint probability of $b_i$ and $f_j$. The definition of conditional probability gives
\begin{equation}
P(b_i, f_j)=P(f_j)P(b_i | f_j)
\end{equation}
where $P(b_i | f_j)$ is defined as
\begin{equation}
P(b_i | f_j)=\frac{a_{ij}}{\omega_j}
\label{omega}\end{equation}
and $\omega_j\equiv\sum_i a_{ij}$ denotes the number of ``synonyms"\cite{origin_Zipf} of $f_j$, i.e., different $b_i$ for the same $f_j$. Note the block itself does not exhibit functions; instead, its collective interactions with environment define the function. Without the information of environment, we assume {\it a priori} probability $P(f_j)=1/q$. 

Now we need to ask whether this assumption is reasonable. Take Ref. \cite{origin_Zipf} for example, it interprets $\mathcal{F}$ as the objects of reference in a language. The fact that $P(f_j)=1/q$ implies the community (environment) regards all objects of reference $f_1, ..., f_q$ are equally important. The above consequence is obviously unreasonable. Via reductio ad absurdum, we rule out the interpretation ``$\mathcal{F}$ is the objects of reference". Let us try another one to test the assumption. As was described in Eqs. (\ref{Book}, \ref{Fr_Book}), $f_j$ denotes ``the function of $s_j$", while the binary matrix ${\bf A}: \mathcal{B}^p\to\mathcal{F}^q$ refers to ``the association matrix". Since the position of $s_j$ in a Book is unique, the probability of finding $f_j$ is $P(f_j)=1/q$ if someone randomly chooses a function in $Fr(\text{Book})$. Now the previous contradiction is resolved. The importance of the position of a block can be exemplified by the following instance: my ``sister" hugs my friend's ``sister". Although we use the same ``sister", their meanings (one kind of communication function) are different due to their distinct position in the context. 

The complexity of a sequence comes from the different blocks it contains. The sequence is preferably to be simple as far as the individual effort is concerned. Such effort is measured by the entropy of block:
\begin{equation}
\mathcal{H}_p(\mathcal{B})=-\sum_{i=1}^{p}P(b_i)\log_p P(b_i).
\end{equation}
When a single block exhibits all functions, namely $P(b_i)=1$, this effort is minimized. Because this effort only considers the production of blocks which is done by the individual, we called it ``individual effort".

In contrast, the sequence tends to be complex from the viewpoint of the collective effort which comes from the fact that, once $b_i$ is produced, the individual will use it to carry out some functions. The collective effort for $b_i$ is defined as
\begin{equation}
\mathcal{H}_q(\mathcal{F} | b_i)=-\sum_{j=1}^{q} P(f_j | b_i)\log_q P(f_j | b_i)
\end{equation}
where $P(f_j | b_i)=P(b_i, f_j)/P(b_i)$; while for $\mathcal{B}$ it is 
\begin{equation}
\mathcal{H}_q(\mathcal{F} | \mathcal{B})=\sum_{i=1}^{p} P(b_i)\mathcal{H}_q(\mathcal{F} | b_i).
\end{equation}
Aware that a function is defined by the collective interaction between blocks and environment, we called $\mathcal{H}_q(\mathcal{F} | \mathcal{B})$ ``collective effort" where the notation $\mathcal{F}|\mathcal{B}$ denotes the individual chose a block before choosing a function.

Combining the individual and collective efforts, the cost function can be defined as Eq. (\ref{tot_effort}). When there are several sequence variations, i.e., changes of ${\bf A}$, we need to compare their $\Omega_{\lambda}({\bf A})$ and select the smallest one. If their efforts are equal, we prefer the variation with a new block because such one has more functions to adapt to the environmental changes. This is so-called the quantified version of natural selection and the principle of least effort. One has to notice the minimization of $\Omega_{\lambda}$ is local, not global - finding out the best ${\bf A}$ to minimize $\Omega_{\lambda}$. Our theory classifies the sequence variation with only two possibilities: the length of Book $q$ will be changed or unchanged. We will elaborate on this in Sec. \ref{FC}.

\subsection{No synonym interpretation}
Before discussing the sequence variation, let us explain how power law is related to the block-function association. The authors of Ref. \cite{origin_Zipf} proposed that Zipf's law comes from the simulation result of Eq. (\ref{P(b_i)}), but they did not mention how to write a Book. If the length of Book equals $L$, the frequency of occurrence for $b_i$ is
\begin{equation}
\rho(b_i)\sim L\times P(b_i)=\frac{L}{q}\sum_j \frac{a_{ij}}{\omega_j}
\end{equation}
where $\omega_j$ is defined in Eq. (\ref{omega}). For a given Book, $\rho(b_i)$ must be an integer. There are two ways to obtain this result. ({\rm i}) The $L$ is large enough to include all possible synonyms. The sum $\sum_j a_{ij}/\omega_j$ can be expressed as a rational number with denominator $D_i$. The requirement that $\rho(b_i)$ must be an integer enforces $L\sim qD_i$. Considering all $b_i$, $L\sim q\Pi_i D_i$ is obviously a number much bigger than a normal Book. Besides, $L\neq q$ is different from our interpretation of $P(f_i)=1/q$ as Eq. (\ref{Book}). So this is infeasible; ({\rm ii})  Follow the interpretation of $P(f_i)=1/q$, $L=q$. The frequency of words in any Book must be an integer, so we necessitate $\omega_j=1$, i.e., ``no synonym", and conclude 
\begin{equation}
\rho(b_i)=\sum_j a_{ij} 
\end{equation}
is an integer. Using simulation as described in Fig. \ref{flowchart}, we can see $\rho(b_i)$ exhibits the power law in Fig. \ref{simulation}. Besides, no synonym interpretation can greatly speed up our evolution algorithm (see SI for detail).

Is ``no synonym" reasonable? The answer is yes. For genetics, if we replace one protein with another similar one, the molecular interaction between environment and protein must change although it may be too small to be detected. For language, similar words are used in slightly different contexts, such as ``because" and ``since". These cases indicate we can state that ``there is no synonym" if we consider $f_j$ in a very rigorous standard. But in the real world, an individual may not be sensitive to the change of blocks with similar functions. Take ``because" and ``since" for instance. If a writer uses 100 ``because" without a single ``since", the reading fluency is greatly reduced. But what about 50: 50 versus 49: 51? The difference is expected to be small. This implies the existence of ``tolerance", the definition of which will be elaborated in Sec. \ref{Tolerance}.

\subsection{Direction of evolution and mutation}\label{FC}
\begin{figure*}[]
	\includegraphics[width=16cm]{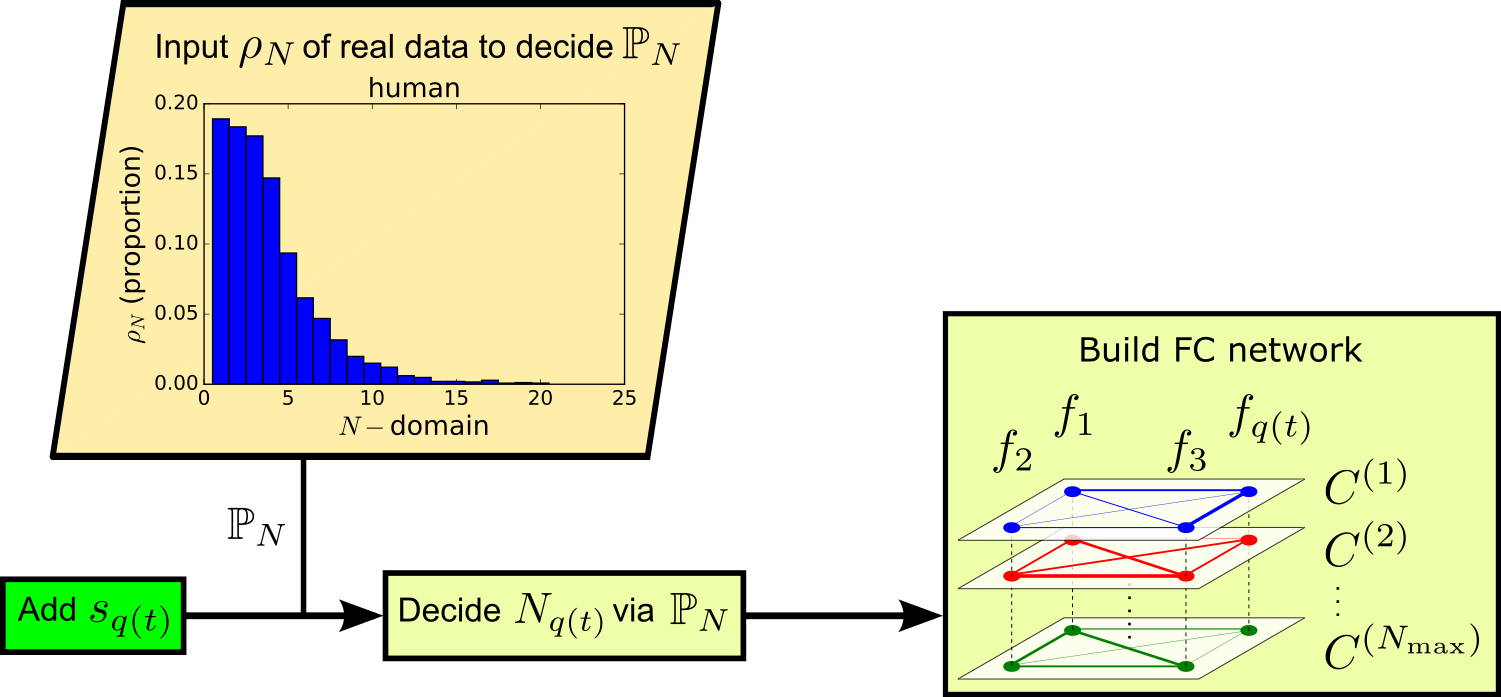}  
	\caption
	{The step ``Add $s_{q(t)}$ and build its FC" in Fig. \ref{flowchart} includes three operations. First, ``Add $s_{q(t)}$": increase the length of Book. Second, ``Decide $N_{q(t)}$ via $\mathbb{P}_N$": estimate the prior probability distribution $\mathbb{P}_N$ from the counts of components in a word $\rho_N$ for real data, then randomly choose $N_{q(t)}$, the number of components in $s_{q(t)}$, on the basis of $\mathbb{P}_N$. Third, ``Build FC network": construct the FC network at the component level for $f_{q(t)}$, where $C^{(k)}$ denotes the FC network for the component at $k-$position and $N_{\max}=\max{\{ N_1, N_2, ..., N_{q(t)}\}}$. Notice that once a block $s_{\mu}$ does not have a component at $k-$position, the FCs between $f_{\mu}$ and all other functions $f_{\nu}$ are set to 0, namely, $C^{(k)}_{\mu, \nu}=C^{(k)}_{\nu, \mu}=0$. 
	In general, $C^{(k)}=C^{(k)}(t)$ is a time-dependent network. It may change whenever $s_{q(t)}$ is added. But when discussing the ``conserved increasing evolution", the above three operations can be modified to accelerate the evolutionary algorithm. See SI for details about our simulation and faster algorithm.
	}\label{FC_network}
\end{figure*}

There are two kinds of sequence variation: change the length of Book $q$ or not. As the evolutionary pressure mounts, the size of Book and $\mathcal{B}$ may change. In our algorithm, we denote the dynamical size of Book as $(p, q)=(p(t), q(t))$ where $t$ is the time step and $(p(0), q(0))=(p_0, q_0)$. In the general case of evolution, both conserved and radical change may happen. It is too complex to analyze all possible evolutionary pathways (See SI for more discussion). But for the simplest case, we can assume ``conserved increase" so that $q(t)=q_0+t$ as in Fig. \ref{flowchart}. To study the sequence variation, we need to construct the FC network. There are two operations as demonstration in Extended Fig. \ref{FC_network}.

The first operation is to decide $N_q{(t)}$ - the number of components of $s_{q(t)}$ in Book. This can be achieved by choosing them based on the prior probability distribution $\mathbb{P}_N$ which can be estimated by measuring real data. For instance, if we want to simulate the FC for a community derived from the following sentence: ``the ham-burg-er con-tains  let-tuce and ba-con", the counts of components in a word $(\rho_1, \rho_2, \rho_3)=(2, 3, 1)$, where $\rho_N$ denotes the number of blocks that contain $N$ components. We then assume the prior probability distribution in this community $(\mathbb{P}_1, \mathbb{P}_2, \mathbb{P}_3)=(2, 3, 1)/6$. See SI for details about the importance of setting a ``reasonable" prior probability distribution.

\begin{figure*}[h!]
	\includegraphics[width=16cm]{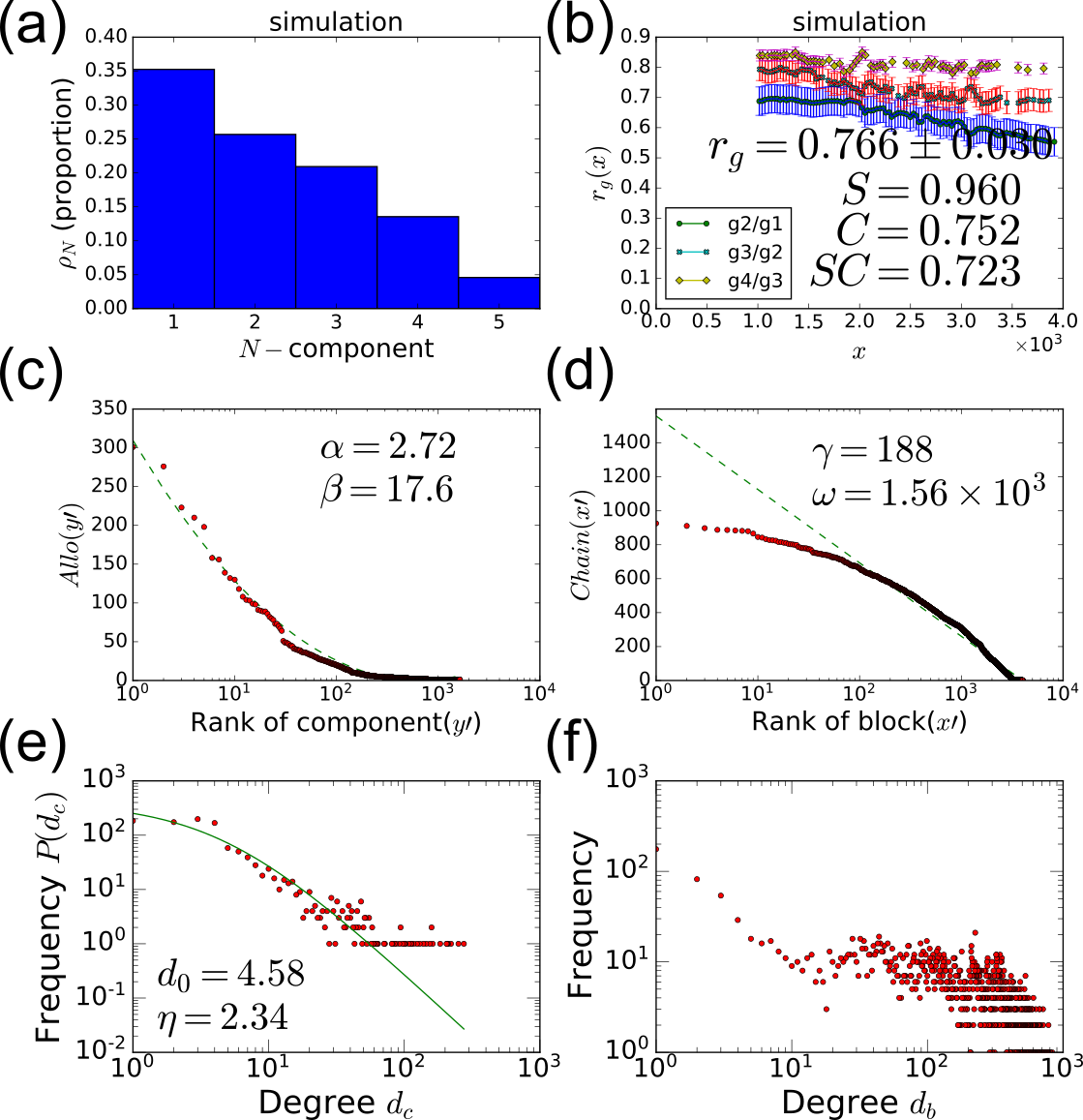}  
	\caption
	{Quantitative characteristics of a simulation sample: (a) Frequency distribution of the number of components in a block; (b) $r_g$,  standard error, and $SC$ value; (c) $Allo$-rank plot; (d) $Chain$-rank plot; Degree distribution of (e) components and (f) blocks. Parameters of this simulation: $L=10000$, $z=0.01L$, $\lambda=0.495$, $P_{\text{mu}}=0.0004$, $T=3$, and the probabilities of creating a 1, 2, 3, 4, 5-component block are set as 0.15, 0.40, 0.25, 0.15, and 0.05, respectively
	}\label{more_simulation}
\end{figure*}

The second operation is to build FC network. Let $C^{(k)}(t)$ be the FC network for the component on $k-$position at time $t$. There are two definitions of position: ({\rm i}) absolute and ({\rm ii}) relative. Let us give an example. For two concepts associated with $s_{\mu}=$``re-cycle" and $s_{\nu}=$``re-flec-tion", ({\rm i}) if we adopt absolute position:  $C^{(1)}_{\mu,\nu}$ denotes the FC between re in $s_{\mu}$ and re in $s_{\nu}$, $C^{(2)}_{\mu,\nu}$ is the FC between cycle in $s_{\mu}$ and flec in $s_{\nu}$, but $C^{(3)}_{\mu,\nu}=0$ because $s_{\mu}$ does not has $3^{rd}$ component; ({\rm ii})  if we adopt relative position: $C^{(begin)}_{\mu,\nu}$ denotes the FC between re in $s_{\mu}$ and re in $s_{\nu}$, $C^{(end)}_{\mu,\nu}$ connects cycle in $s_{\mu}$ and tion in $s_{\nu}$, but $C^{(mid)}_{\mu,\nu}=0$ since $s_{\mu}$ only has begin and end components. Note in both ({\rm i}) and ({\rm ii}), $C^{(k)}_{\mu, \nu}(t)=0$ once $s_{\mu}$ or $s_{\nu}$ does not have the $k-$position component. To present the simplest result of function connection model, we assume $C^{(k)}_{\mu, \nu}(t)=C^{(k)}_{\nu, \mu}(t)$ are random numbers between 0 and 1. Although definition ({\rm ii}) seems often in the real world, it is hard to define precisely. Thus, we shall adopt a definition ({\rm i}) that can equally reproduce the features in GLC (see Fig. \ref{simulation} and \ref{more_simulation}), but is comparatively easier to handle. 

Now we need to determine the composition of $s_{q(t)}$. There are two modes ({\rm I}) co-option and ({\rm II})  {\it de novo}. 

In ({\rm I}) co-option, $f_{q(t)}$ will be associated with an old block $s_j$ on Book  where $1\le j\le q(t-1)$. The probability that $f_{q(t)}$ uses $s_j$ is:
\begin{equation}
P^{old}_{q, j}(t)=\frac{C_{q, j}(t)}{\sum_{\mu} C_{q, \mu}(t)}
\label{P_old}\end{equation}
where $C_{q, \mu}(t)\equiv\sum_{k=1}^{d(t)} C^{(k)}_{q, \mu}(t)$, $d(t)=\min(N^{[t]}_q, N^{[t]}_\mu)$, $\forall 1\le \mu \le q(t-1)$. If a certain block $s_\zeta$ is selected, the association matrix ${\bf A}$ will require new elements
\begin{equation}
a_{i, q(t)}= \begin{cases}
1,	&\text{if } b_i=s_\zeta\\
0,	&\text{otherwise}
\end{cases}
\label{old_A}\end{equation}
where $1\le i\le p(t-1)=p(t)$. The size of $\mathcal{B}$ does not change.

In ({\rm II}) {\it de novo}, we need to consider the situation that $f_{q(t)}$ uses new components. For genetics, it may come from other organisms (transduction or conjugation), environments (transformation), or even non-coding DNA ({\it de novo} gene birth). For linguistics, it may come from other languages, sounds in nature, or even a sound that has never been used. To describe this, we can define a new quantity - effective connection $z$. The probability of ``creating" the $k^{th}$ component for $f_{q(t)}$ is:
\begin{equation}
P^{(k)}_{q, new}(t)=\frac{z}{z+\sum_{\mu} C^{(k)}_{q, \mu}(t)}.
\end{equation}

We can also recombine existing components to form a new block $s_{q(t)}$ for $f_{q(t)}$. The probability that $f_{q(t)}$ uses a $k^{th}$ component in $s_j$ is:
\begin{equation}
P^{(k)}_{q, j}(t)=\frac{C^{(k)}_{q, j}(t)}{z+\sum_{\mu} C^{(k)}_{q, \mu}(t)}
\end{equation}
where $P^{(k)}_{q, new}(t)+\sum_{j=1}^{q(t-1)} P^{(k)}_{q, j}(t)=1$. The components at position $k$ for $s_{q(t)}$ will be created on the basis of $P^{(k)}$ where $k=1, 2, ..., N_{q(t)}$. The association matrix ${\bf A}$ will require additional elements \begin{equation}
a_{p(t), j}= \begin{cases}
1,	&\text{if } j = q(t)\\
0,	&\text{otherwise}
\end{cases}
\label{new_A}\end{equation}
and $a_{i, q(t)}=0\ \forall 1\le i\le p(t-1)=p(t)-1$. The size of $\mathcal{B}$ increases by one. In fact, there is a small probability that we ``create" an already existing block $b_j$ for $s_{q(t)}$. If so, ${\bf A}$ will be changed according to Eq. (\ref{old_A}) instead of Eq. (\ref{new_A}). Since both ({\rm I}) co-option and ({\rm II}) {\it de novo} modify $\mathcal{B}^p\to\mathcal{F}^q$, the use of Eq. (\ref{tot_effort}) for Fig. \ref{flowchart} will decide which mode is better. The above procedures lay out the recipe for function connection in sequence variation.

For the variation that does not affect $q(t)$, we incorporate it with mutation, that is, ${\bf A}$ changes by chance. For instance, the usage of ``flyer" has been replaced by ``airplane", and ``thou" by ``you". To quantify this feature, one can simply assign each $f_j where 1\le j\le q(t)$ a probability of occurrence $P_{\text{mu}}$ for mutation. When mutation happens, there are three situations (see SI for detail): no change, a certain block is replaced by an existing one (co-option like mutation), and create a block to substitute the original one ({\it de novo} like mutation). To decide which case survives, we again employ the principle of least effort, as in Fig. \ref{flowchart}. 

In real evolution, mutation may repeat during a certain time step $t$. We introduce the repeat count of mutation $\tau$ to simulate this feature. After our algorithm completes execution for mutation of Book (from $j = 1 \sim q$), the repeat count of mutation $\tau$ will increase by one. By setting a maximum repeat count $T \ge \tau$, we can control the amount of mutation.

Our simulation exhibits the features of GLC for both life and language by adjusting $P_{\text{mu}}$ and $z$. A higher $P_{\text{mu}}$ renders the FRD of block $\rho_x$ more resembling that of language, while a higher $z$ causes the FRD of component $\rho_y$ more curved. This result is consistent with the fact that evolution of language is much faster than that of life.

\subsection{Tolerance of synonym}\label{Tolerance}
Now let us go back to the ``no synonym" assumption. Having seen that $f_j$ is related to FCs, we can define synonym in the usual sense as ``if replacing $b_l$ at $s_j$ by another block $b_q$ has little influence on its FCs, we accept $b_q$ as the synonym of $b_l$ at $s_j$". Mathematically, we can write down the change of FC for $f_j$ as
\begin{equation}
\Delta f_j(b_q|b_l)\equiv \frac{\sum_\alpha \Big(C_{q, \alpha}-C_{l, \alpha}\Big)^2}{\sum_\beta C_{l, \beta}^2}.
\end{equation}
The definition of synonym is thus equivalent to requiring $\Delta f_j(b_q|b_l)$ smaller than tolerance $T_f$. It will be worthwhile to determine the value of $T_f$ that depends on the interactions between individual and environment.

\end{document}